\documentclass[]{article}
\usepackage[a4paper, margin=2.5cm]{geometry}

\usepackage{amsmath,amssymb,amsfonts}%

\usepackage{microtype}
\usepackage{polyglossia}
\setmainlanguage[]{english}

\usepackage[font={small},labelfont={bf}, labelformat=simple, labelsep=period, format=hang]{caption}

\usepackage{graphicx}%

\usepackage{setspace}
\usepackage[style=nature, maxbibnames=99, doi=true, isbn=false, url=false, eprint=false, date=year, sorting=none, backend=biber]{biblatex}
\AtEveryBibitem{\clearlist{language}}
\usepackage{hyperref}
\hypersetup{pdfpagemode=UseNone}

\begin{document}
\newgeometry{left=3.1cm, bottom=3cm, right=3.1cm, top=3cm} 
\setlength{\parindent}{0cm}

\begin{center}
    {\Large \textbf{Quantum-Well-Metasurface to Maximize Nonlinear Polarization}\par \bigskip}
    Pernille Undrum Fathi\textsuperscript{1, *},
    Irene Occhiodori\textsuperscript{1, 2},
    Patrick Devaney\textsuperscript{3},
    Amberly Ricks\textsuperscript{3},\\
    Rithvik Ramesh\textsuperscript{3, 4},
    Yiwei Ju\textsuperscript{5},
    Moaz Waqar\textsuperscript{5},
    Theodore P. Letsou\textsuperscript{1, 6},\\
    Christina M. Spägele\textsuperscript{1},
    Hyunseung Jung\textsuperscript{7},
    Igal Brener\textsuperscript{7},
    Xiaoqing Pan\textsuperscript{5, 8, 9},\\
    Marcus Ossiander\textsuperscript{1, 10, *},
    Seth R. Bank\textsuperscript{3},
    Federico Capasso\textsuperscript{1, *}
    \bigskip
\end{center}

{\footnotesize
\textsuperscript{1} Harvard John A. Paulson School of Engineering and Applied Sciences, Harvard University, Cambridge, 02138, Massachusets, USA

\textsuperscript{2} Department of Materials, ETH Zürich, Zürich, 8093, Switzerland

\textsuperscript{3} Microelectronics Research Center and Electrical and Computer Engineering Department, The University of Texas at Austin, Austin, 78712, Texas, USA

\textsuperscript{4} Department of Electrical Engineering, California Institute of Technology, Pasadena, 91125, California, USA

\textsuperscript{5} Department of Materials Science and Engineering, University of California, Irvine,  Irvine, 92697, California, USA

\textsuperscript{6} Department of Electrical Engineering and Computer Science, Massachusetts Institute of Technology, Cambridge, 02142, Massachusets, USA

\textsuperscript{7} Center for Integrated Nanotechnologies, Sandia National Laboratories, Albuquerque, 87185, New Mexico, USA

\textsuperscript{8} Department of Physics and Astronomy, University of California, Irvine,  Irvine, 92697, California, USA

\textsuperscript{9} Irvine Materials Research Institute (IMRI), University of California, Irvine,  Irvine, 92697, California, USA

\textsuperscript{10} Institute of Experimental Physics, Graz University of Technology, Graz, 8010, Austria}
\bigskip

* pernillefathi@g.harvard.edu, marcus.ossiander@tugraz.at, capasso@seas.harvard.edu
\bigskip
\bigskip

\textbf{
Nonlinear frequency conversion unlocks technologies ranging from telecommunications to quantum computation; however, weak nonlinearities and architectures that resist miniaturization currently limit devices. Here, we combine a bandstructure-engineered GaAs/AlGaAs heterostructure with a high quality factor dielectric metasurface to simultaneously tailor the intrinsic nonlinear susceptibility and optimize the electromagnetic field within the heterostructure. By engineering a resonant interband transition, we realize a large second-order nonlinear tensor element, 1.6 nm/V at 1.57 $\boldsymbol{\mu}$m wavelength. We then make it free-space-accessible and boost the effective nonlinearity to $\boldsymbol{\sim}$ 14 nm/V using a metasurface patterned on the material. Our proof-of-concept experiment establishes that interband transition engineering and metasurfaces accessing otherwise unusable nonlinear tensor elements enable giant effective nonlinearities in the near-infrared to visible spectrum. This addresses material and device-level constraints in nonlinear photonics, providing a scalable route to compact, efficient devices.}

\restoregeometry 
\setlength{\parindent}{0.5cm}
\newpage
Nonlinear optical processes are fundamental to modern photonics, enabling frequency mixing \cite{liu_all-dielectric_2018}, ultrafast signal processing \cite{wu_advancements_2025}, broadband and pulsed light sources \cite{saghaei_broadband_2019, jankowski_supercontinuum_2023, chang_integrated_2022}, high-speed modulation \cite{pan_ultra-compact_2022}, and the generation of entangled photon pairs for quantum communication 
\cite{noh_quantum_2024, santiago-cruz_resonant_2022, zhang_spontaneous_2021}. 
Nonlinear processes arise when the electric field of intense light drives the electrons in matter far enough from their equilibrium positions that the restoring force is no longer proportional to their displacement. The electrons' motion becomes anharmonic, and the resulting material polarization becomes nonlinear in the applied electric field strength, $E(t)$. Well below the damage threshold of a material, the polarization's power series expansion $P(t) = P_L(t) + P_{NL}(t) = P_L(t) + \epsilon _0 \left( \chi^{(2)}E^2(t) + \chi^{(3)}E^3(t) +\dots \right)$ converges and allows separating the linear $P_\text{L}$ and nonlinear $P_\text{NL}$ contributions. The latter allows for creation of frequencies not present in the incident light and exchanging information and energy between photons. However, it is weak compared to the linear material polarization response (for now, we disregard effects of the light polarization and the frequency dependence of the nonlinear susceptibilities $\chi^{(2)}, \chi^{(3)}, \dots$).

Efficient use of nonlinear effects requires light with high electric field strengths and materials with large nonlinear susceptibilities. Conventional nonlinear devices rely on a handful of bulk crystals with weak intrinsic nonlinearities. To compensate for that, these systems typically use long propagation distances to build up nonlinearly generated radiation coherently \cite{yang_symmetric_2024, wang_second_2017}. This approach hinders device miniaturization and requires phase-matched propagation conditions, which necessitate complex dispersion engineering.

Metasurfaces are subwavelength spaced nanoscatterers, known for their versatility in tailoring the properties of electromagnetic waves \cite{pendry_controlling_2006, yu_light_2011}. This versatility enables the miniaturization of classical optical components with so-called flat optics \cite{yu_flat_2014, chen_flat_2020}, novel active and tunable components \cite{jung_rise_2024, abdelraouf_recent_2022}, and applications in quantum optics \cite{noh_quantum_2024, santiago-cruz_resonant_2022}. For applications in nonlinear photonics, metasurfaces can enhance incident light fields and thereby enable efficient nonlinear devices with short propagation distances \cite{Lee2014, carletti_enhanced_2015, gili_monolithic_2016, camacho-morales_nonlinear_2016, li_nonlinear_2017,krasnok_nonlinear_2018, fedotova_second-harmonic_2020}. Still, the nonlinear susceptibilities of the underlying material platforms fundamentally limit their performance. 

Semiconductor multi-quantum-well (MQW) heterostructures overcome the limited nonlinear response of bulk crystals. Bandstructure engineering can create MQW states and control the mean position of an electron in these states. An intersubband transition between two such states with a large spatial offset is then associated with a strong electron displacement, which can resonantly enhance the nonlinear susceptibilities close to the intersubband transition energy \cite{kroemer_nobel_2001, esaki_superlattice_1970}. However, in semiconductors, intersubband transition energies are limited by the band offset of the heterostructure. This fact restricts operating wavelengths of III-V semiconductor heterostructures to the mid-infrared, where this approach to tailored nonlinear response has been studied and heterostructures can boost second-order nonlinear susceptibilities by orders of magnitude compared to bulk crystals \cite{fejer_observation_1989, sirtori_observation_1991, capasso_coupled_1994}. Field enhancement in plasmonic and Mie resonators can improve mid-infrared nonlinear conversion efficiencies \cite{Lee2014, Sarma2022, mann_inverse_2023}. The use of materials with wide band offsets can enable intersubband engineering at near-infrared wavelengths \cite{quach_ganaln_2020}, but demonstrated nonlinear heterostructures have been restricted to materials with loss at the pump and the harmonic wavelengths \cite{qian_large_2019, chen_giant_2024}.

To overcome the spectral constraint of intersubband transitions in semiconductors while maintaining the ability to tailor the nonlinear material response, Khurgin proposed using transitions between MQW states located in different bands, i.e., interband transitions \cite{khurgin_secondorder_1987, khurgin_second-order_1988}. To distinguish these from the interband transitions present in bulk crystals, we will refer to them as MQW-interband transitions.

Recent studies numerically predicted a MQW-interband-transition-engineered GaAs/AlGaAs heterostructure with a $\chi^{(2)}$ that exceeds that of bulk GaAs by an order of magnitude \cite{Ramesh2023}. In 2025, a material with a MQW-interband-transition-enhanced $\chi^{(2)}$ was demonstrated \cite{ramesh_enhanced_nodate}, albeit using a nonlinear tensor element that requires an excitation polarization scheme which is inefficient in free-space applications. 

Here, we use designed MQW-interband transitions combined with dielectric metasurface field enhancement to realize accessible, giant effective nonlinearities for near-infrared-to-visible light conversion. The MQWs provide one of the largest second-order nonlinear tensor elements in this spectral region, and the metasurface converts it to an accessible and further enhanced effective tensor element.

\begin{figure}[tb]
    \centering
    \includegraphics[width=0.95\textwidth]{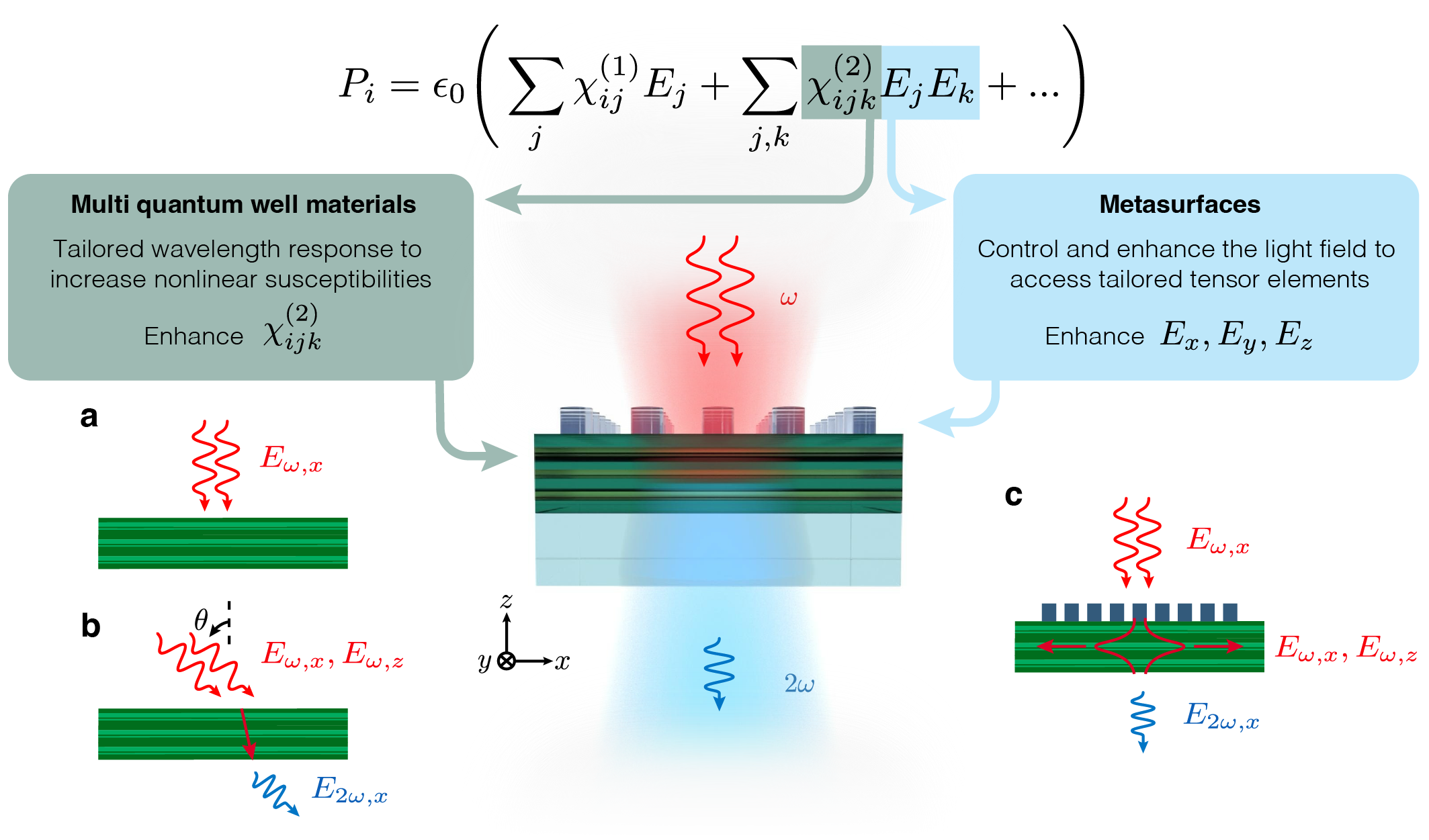}
    \caption{ \textbf{Tailoring all components of the second-order polarization.}
    Combining asymmetric multi-quantum-wells, which enable resonance enhancement of higher-order susceptibilities, with metasurfaces, which enable precise control and enhancement of electromagnetic fields, enables devices in which every factor of a nonlinear polarization term is tailored to the application. 
    \textbf{a} At normal incidence, selection rules prohibit second harmonic generation enhanced by MQW-interband transitions.
    \textbf{b} At oblique incidence, selection rules are fulfilled and second harmonic generation is possible. For free space excitation, the conversion efficiency is limited by the maximum propagation angle achievable within the high index medium. Note that this material requires two different polarizations for the pump photons. 
    \textbf{c} Metasurfaces enable polarization conversion and resonant field enhancement, thereby enabling enhanced second harmonic generation.
    }
    \label{fig:concept}
\end{figure}

\section*{Aspects of Nonlinear Frequency Conversion}

Figure \ref{fig:concept} illustrates our concept. Generally, the material polarization $\vec{P}$ depends on the incoming light polarization, i.e., the direction of the light's electric field vector $\vec{E}$. In non-isotropic materials, these vectors are not necessarily aligned, requiring a second-rank tensor with the elements $\chi^{(1)}_{ij}$ to describe the linear material response (where $i$ denotes a material polarization direction and $j$ denotes an electric field direction), a third-rank tensor with the elements $\chi^{(2)}_{ijk}$ to describe the second-order nonlinear response, and so on. In the following, we focus on the second-order nonlinear response applied to the special case of second harmonic generation, in which two incident similar-frequency photons are annihilated to create one photon containing their total energy.

When assuming a dominant $\chi^{(2)}_{ijk}$ element, maximizing the second-order nonlinear material response along a direction $i$ in response to a given incident light field strength polarized along the $j$ and $k$ direction requires a large $\chi^{(2)}_{ijk}$, and comparable electric field strengths $E_j \approx E_k$.

To achieve the first, we use a heterostructure with a MQW-interband-transition-enhanced second-order nonlinear susceptibility. The heterostructure size is limited to micrometers along the growth direction, which we denote as $z$, and consists of quasi-infinite GaAs/AlGaAs layers along the $x$ and $y$ directions. Due to its $x$-$y$ translational symmetry, the heterostructure can only control the spatial position of states, and thus electron motion, along the growth direction $z$. Selection rules designate $\chi^{(2)}_{xzx}, \chi^{(2)}_{xxz},\chi^{(2)}_{yzy},\chi^{(2)}_{yyz}$ as the dominant tensor elements (see below). 

This poses a challenge for free space applications, see Fig. \ref{fig:concept}a: light must possess a polarization component along the growth direction to exploit the enhanced tensor elements, i.e., it cannot be incident along the growth direction. Illuminating the material with light perpendicular to the growth direction provides the correct polarization components, but restricts the coupling area to a small illuminated facet, such that achieving a total nonlinear polarization requires substantially higher optical intensities. In practice, the usable optical power is therefore limited by the material damage threshold. The remaining option to create an electric field along the growth direction is oblique incidence (Fig. \ref{fig:concept}b). However, even at Brewster's angle, which prevents reflection losses, the high refractive indices of III-V semiconductors reduce the light's internal propagation angle and the field strength along the growth direction. 

We overcome this challenge by implementing a metasurface that generates an electric field component along the heterostructure's growth direction from normally incident light using a guided mode resonance, see Fig. \ref{fig:concept}c and below for details \cite{fan_guided_1995, huang_ultrahigh-q_2023, muhowski_mid-infrared_2022, bark_transmission_2018}. At the same time, the resonance confines light to the heterostructure, enhancing the in- and out-of-plane electric field and consequently the nonlinear material polarization.

In summary, we exploit MQW-interband transitions in heterostructures to enhance select nonlinear tensor elements to values far beyond those found in naturally occurring materials. We then convert the incident light polarization using a dielectric metasurface to access these nonlinear tensor elements and confine the incident electric field to the heterostructure. This simultaneously enhances all factors contributing to the second-order-nonlinear polarization product (see Fig. \ref{fig:concept}). The result is an effective second-order nonlinear susceptibility behaving like the $\chi^{(2)}_{xxx}$ of a bulk crystal. We demonstrate, to our knowledge, a record magnitude at visible and near-infrared frequencies of $\chi^{(2)}_{\mathrm{eff,MQW+MS}} = \chi^{(2)}_{xxx\mathrm{,MQW+MS}} \approx 14$ nm/V.


\section*{Material Design and Characterization}

\begin{figure}[thb]
    \centering
    \includegraphics[width=0.95\textwidth]{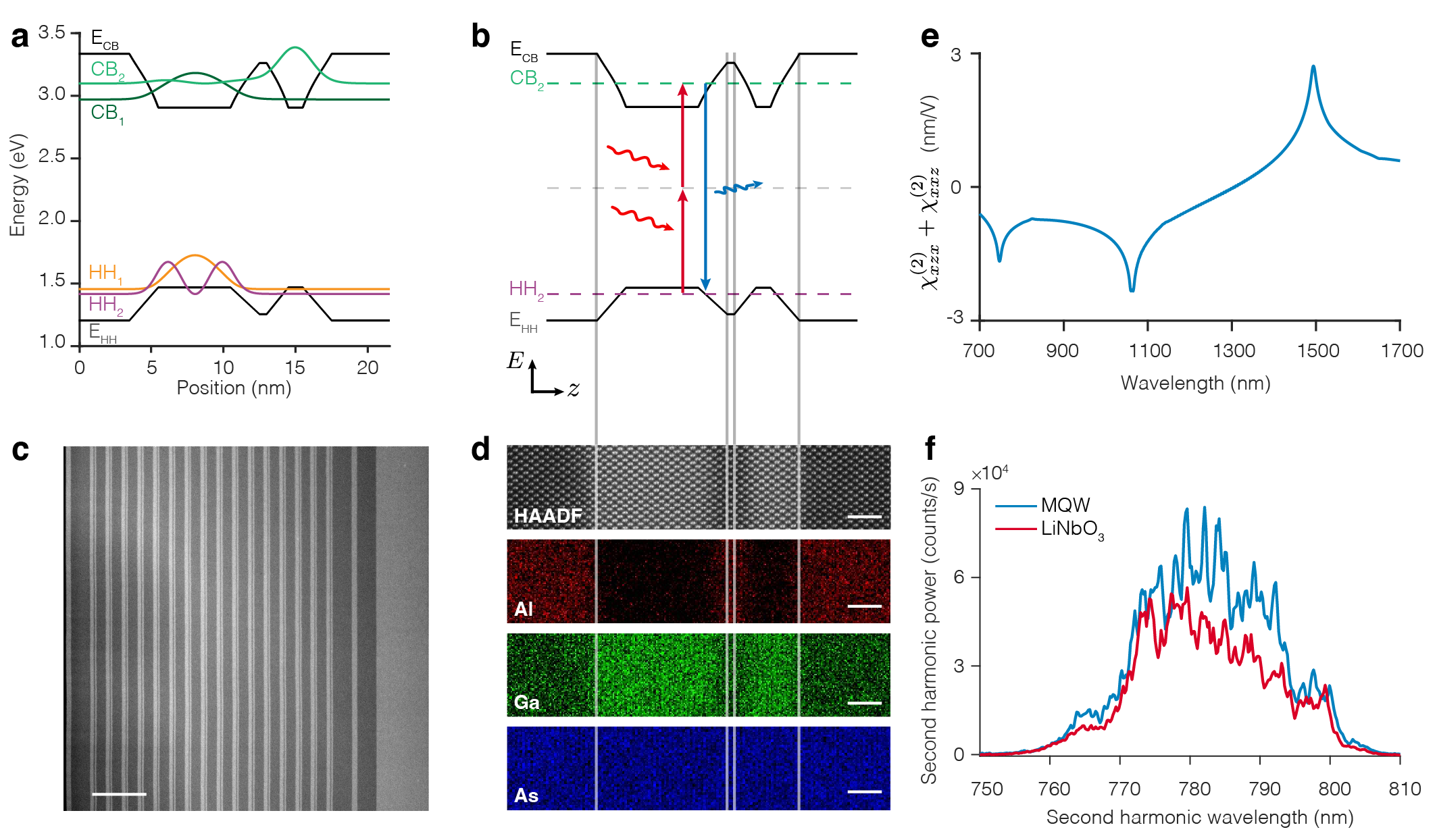}
    \caption{ \textbf{GaAs/AlGaAs heterostructure properties.}
    \textbf{a} Band-edge diagram of a single GaAs/AlGaAs double well showing the energies (E$_{HH}$, E$_{CB}$) and the probability distributions of the first two heavy hole (HH$_1$, HH$_2$) and conduction band (CB$_1$, CB$_2$) states. The diagram shows that the heavy hole states and the first conduction band state are confined to the wider quantum well, while the second conduction band state is confined to the thinner well. 
    \textbf{b} Band-edge diagram demonstrating the transitions involved in the HH$_2$ to CB$_2$ nonlinear enhancement. The asymmetry of the structure leads to an offset between the centroids of the probability distributions of the two states, giving rise to a dipole-like diagonal intersubband matrix element creating a large second-order nonlinearity.
    \textbf{c} High-angle annular dark-field scanning transmission electron microscopy (HAADF-STEM)  image of a heterostructure cross-section showing multiple GaAs/AlGaAs double wells (scale bar: 100 nm). 
    \textbf{d} High-magnification HAADF-STEM image (top) and corresponding energy-dispersive X-ray spectroscopy (EDS) maps of Al, Ga, and As in a single period of the MQW structure  (scale bar: 2 nm).  
    \textbf{e} Simulated second-order susceptibility of the structure in \textbf{a} as a function of the pump frequency during second harmonic generation (smoothed using a third-order Savitzky–Golay filter).
    \textbf{f} Experimental second harmonic spectrum generated by the MQW measured at $45^{\circ}$ incidence angle and x-cut LiNbO$_3$ at $0^{\circ}$ incidence angle (after subtracting a constant background intensity). 
    }
    \label{fig:material}
\end{figure}

Inversion symmetric materials cannot generate second harmonic radiation because the material polarization induced by the positive and negative half-cycles of incoming light cannot differ. Conversely, a strong spatial asymmetry, i.e., an environment that favors electron motion in one but not the opposite direction of motion, is a prerequisite for a large second-order susceptibility. 

To induce such asymmetry, we couple two GaAs/AlGaAs quantum wells with different widths to form the basis of our heterostructure. Figure \ref{fig:material}a shows their band diagram and the probability distributions of the first two heavy hole and conduction band states. The second heavy hole state (HH$_2$) is confined in the wider of the two quantum wells, whereas the second conduction band state (CB$_2$) is confined in the thinner well \cite{Ramesh2023}. 
The HH$_2$ to CB$_2$ interband transition is therefore associated with a large electron displacement (towards the right in Fig. \ref{fig:material}a) without an opposite-displacement counterpart. This asymmetry substantially enhances the second-order nonlinear properties of the double-well close to the transition energy between the HH$_2$ and CB$_2$ states when the selection rules are satisfied. 

Interband transitions involving a heavy hole state require a photon polarized perpendicular to the growth direction (along $x, y$) of the layers to satisfy the interband selection rules at the $\Gamma$ point \cite{oppo_optical_2020}. One pump photon and the second harmonic photon share this polarization. To benefit from the tailored asymmetry, which the double well only enforces along the growth direction, the second pump photon must be polarized along the growth direction ($z$) \cite{faist_quantum_2013}. 

Generally, $\chi^{(2)}_{ijk}(\omega_3, \omega_1, \omega_2)$ depends on the frequencies of three involved photons; however, choosing two photon frequencies determines the third photon's frequency via energy conservation, i.e., $\omega_3 = \omega_1 + \omega_2$ (see Fig. \ref{fig:material}b). In transparent bulk crystals, when all frequencies are far from resonance (i.e., when absorption and dispersion are negligible), symmetries reduce the number of independent $\chi^{(2)}_{ijk}(\omega_1+\omega_2, \omega_1, \omega_2)$ tensor elements. As our heterostructure uses resonant enhancement, this approach is inapplicable. However, the second-order nonlinear susceptibility responsible for second harmonic generation only depends on the frequency of the incident photons $\chi^{(2)}_{ijk}(\omega) = \chi^{(2)}_{ijk}(2\omega, \omega, \omega)$. In this case, the choice of $\chi_{ijk}(\omega)$ is ambiguous because the $\chi^{(2)}_{\text{eff, MQW}}(\omega)$ relevant for second harmonic generation remains unchanged as long as $\chi^{(2)}_{ijk}(\omega)+\chi^{(2)}_{ikj}(\omega)$ remains constant \cite{boyd_nonlinear_2019}. For clarity, we enforce intrinsic permutation symmetry for the indices $j, k$, yielding $\chi^{(2)}_{ijk}(\omega)=\chi^{(2)}_{ikj}(\omega)$ in the following. Note that the equality is untrue for sum-frequency generation, i.e., when $\omega_1 \neq \omega_2$. 

At the transition energy, the selection rules for the MQW-interband transition yield dominant $\chi^{(2)}_{xzx}= \chi^{(2)}_{xxz}= -\chi^{(2)}_{yzy} = -\chi^{(2)}_{yyz}$ tensor elements with large magnitudes and opposite signs in the near-infrared.

To amplify $\chi^{(2)}_{xzx}$, we adjust the barrier thickness, degree of asymmetry, and overall size of the double quantum well. These parameters tune the overlap and centroids of the HH$_2$ and CB$_2$ probability distributions. For details on the design, growth and characterization, see Methods.

For our demonstration, we use a 16-period MQW film with a total thickness of $595$ nm, bonded to a sapphire substrate with epoxy resin. Neighboring double-wells are separated by a thick AlGaAs barrier to prevent adjacent quantum well states from hybridizing. Figure \ref{fig:material}c shows a cross-section High-angle annular dark-field scanning transmission electron microscopy (HAADF-STEM) image of the grown heterostructure. Fig.~\ref{fig:material}d shows the constituent atom concentration in one double well, determined via energy-dispersive X-ray spectroscopy (EDS) (see Methods). The data shows that epitaxial growth leads to imperfect interfaces that, unlike the typical depiction of a quantum well, lead to a band slope at the GaAs/AlGaAs interfaces. Using this experimentally corrected band profile enables numerical simulations of the corresponding, realistic nonlinear response. Fig. \ref{fig:material}e shows the computed spectrally resolved $\chi^{(2)}_{xzx}(\omega)+\chi^{(2)}_{xxz}(\omega)$. Methods and Refs. \cite{Ramesh2023, ramesh_enhanced_nodate} describe the simulations. 

We illuminate the heterostructure with $x$-polarized, 70-fs long laser pulses (see the Supplementary Information) to determine its $\left|\chi^{(2)}_{xzx}+\chi^{(2)}_{xxz}\right|$ in an optical measurement, and compare the relative second harmonic flux generated by our sample (at $45^{\circ}$ incidence angle) with that generated by an x-cut LiNbO$_3$ thin-film (at $0^{\circ}$ incidence angle, the $z$-axis of the LiNbO$_3$ aligns with our laser pulses' polarization) under otherwise equal experimental conditions. Fig.~\ref{fig:material}f compares the spectrally resolved second harmonic fluxes. Calibrating the measurement with the known $\chi^{(2)}_{zzz, \text{ LiNbO$_3$}}=51.9 $ pm/V of LiNbO$_3$ (see Methods), the refractive indices of the samples, and the reflection, absorption, and thin-film interference in our samples yields $\chi^{(2)}_{\text{eff, MQW}} = \left|\chi^{(2)}_{xzx}+\chi^{(2)}_{xxz}\right|= 1.6$ nm/V at 1.57 $\mu$m, four times that of bulk GaAs ($\chi^{(2)}_{xyz, \text{ GaAs}}$ = 370 pm/V \cite{boyd_nonlinear_2019}) and two orders of magnitude larger than that of LiNbO$_3$. Not accounting for the absorption of the second harmonic radiation in the heterostructure slightly reduces the usable $\chi^{(2)}_{\text{eff, MQW}}$ to $1.2$ nm/V at 1.57 $\mu$m.
Notably, despite its large $\chi^{(2)}_{\text{eff, MQW}}$, the bare MQW film only produces second harmonic powers comparable to the similarly thick LiNbO$_3$ film, highlighting the need for a different excitation scheme in the case of MQW-interband-driven nonlinearities.

The measured $\left|\chi^{(2)}_{\text{eff, MQW}}\right|$ reaches $\sim 60\%$ of the numerical prediction shown in Fig.~\ref{fig:material}e. This difference is explained by the relatively thick AlGaAs barriers separating subsequent double wells, which do not contribute to the nonlinearity enhancement, lowering the average response of the thin-film. Even higher experimental $\chi^{(2)}_{\text{eff, MQW}}$ can be achieved by reducing this barrier thickness, or by improving the MQW interface sharpness through the introduction of growth interruptions during molecular beam epitaxy \cite{ramesh_enhanced_nodate}. 

\section*{Metasurface Design}

\begin{figure}[thb!]
    \centering
    \includegraphics[width=0.95\textwidth]{figures/device.png}
    \caption{ \textbf{Metasurface design and device simulations.}
    \textbf{a} Metasurface unit cell with dimensions (dark blue: TiO$_2$, green: heterostructure, light blue: Al$_2$O$_3$). 
    \textbf{b} Field distribution in the waveguide at resonance. The panels show the spatial ($x$-$y$) profile of the simulated fields (arb.u.) in the waveguide, summed along the growth direction ($z$). We show the $x$ and $z$ component of the pump field and the $x$ component of the second harmonic field for normal and $0.3^\circ$ incidence angle for the left branch of resonance  B (see below). For normal incidence, the symmetry of the fields yields a vanishing $\beta$.
    \textbf{c} Simulated transmission spectrum of the sample for $x$-polarized light incident at angles between $0^\circ$ and $8^\circ$, introduced by tilting the sample about the $y$-axis. We observe three GMRs labeled A, B, and C. Resonances A and B split spectrally for finite incidence angles. Resonance C requires a finite incidence angle from tilting about the $x$-axis to spectrally split.
    \textbf{d} Simulated modal overlap spectrum ($\beta (\lambda_0$), where $\lambda_0$ is the vacuum wavelength of the pump) of the sample for $x$-polarized light incident at angles between $0^\circ$ and $8^\circ$, demonstrating negligible overlap at normal incidence.
    At finite incidence angles, the breaking of the electric field  symmetry (see panel b) enables large peaks in the overlap that closely follow the spectral positions of the GMRs.
    \textbf{e} Simulated transmission spectrum for $x$-polarized light incident at $0.3^\circ$, corresponding to the cut marked in \textbf{c}.
    \textbf{f} Simulated modal overlap spectrum for $x$-polarized light incident at $0.3^\circ$, corresponding to the cut marked in \textbf{d}.
    }
    \label{fig:device}
\end{figure}

To access the $\chi^{(2)}_{xzx}$ and $\chi^{(2)}_{xxz}$ tensor elements of the MQW film, we employ metasurface-enabled resonances to create the necessary light polarization components and enhance the electromagnetic field in the heterostructure. Metasurfaces can sustain a variety of resonant responses, from localized Mie modes of individual nanostructures to collective, nonlocal effects arising from periodic arrays of scatterers. Here, we use guided mode resonances (GMRs), a nonlocal resonance type \cite{fan_guided_1995, huang_ultrahigh-q_2023, muhowski_mid-infrared_2022, bark_transmission_2018}:

Conceptually, GMRs can be understood in terms of a grating coupler to a waveguide: the grating, placed on an extended, two-dimensional, high refractive index layer, diffracts incident light. If the diffracted propagation angle of the light allows total internal reflection in the layer, the light is coupled in and propagates along the guiding layer until it is coupled out (often by the same grating). Efficient coupling is achieved in the case of constructive interference in the waveguide, i.e., if the propagation phase accumulated between any two points in the layer matches the phase difference of the light incident on these two points. This condition yields sharp resonances if the grating size (period number) is large and the coupling strength is weak. Because the resonant modes propagate along the waveguide, they can create strong $E_x$ and $E_z$ fields, which can in turn drive the $\chi^{(2)}_{xzx}, \chi^{(2)}_{xxz}$ nonlinearity. 

To achieve such a GMR at a pump wavelength of 1.58 $\mu$m, we deposit a TiO$_2$ nanopillar array, acting as the sub-wavelength diffractive grating, on top of the heterostructure, which acts as the waveguide. The nanopillar array matches the momentum of the incident light to that of a mode guided in the heterostructure. First, we design a GMR that enables the correct light polarization components $E_{x, \omega}, E_{z, \omega}$ and a high quality factor, thereby long resonance lifetime, and strong field enhancement, by tuning the metasurface's periodicities, height, and pillar dimensions. To ensure efficient second harmonic generation, we then search for a high modal overlap factor, $\beta$, between the pump and the second harmonic fields $E_{x, 2\omega}$ in the heterostructure \cite{roke_nonlinear_2004, rose_circular_2013, carletti_enhanced_2015, noor_mode-matching_2020}.

\begin{equation}
    \beta \propto \Big | \int \limits_{\text{MQW}} E_{x, \omega}E_{z, \omega} E^*_{x, 2\omega} d^3x \Big |
\end{equation}

Fig. \ref{fig:device}a illustrates the resulting device unit cell. Fig. \ref{fig:device}c,e shows its spectrally-resolved transmission for incidence angles between $0 ^\circ$ and $8 ^\circ$ as the sample is rotated about the $y$-axis, and Fig.~\ref{fig:device}d, f shows the corresponding modal overlap. The displayed spectral range contains three GMRs, labeled A, B, and C.

An inversion symmetric grating imparts equal momentum, $m\vec{k}_g$, in the backward and forward spatial direction, where $m$ is the integer grating order. Thus, at normal incidence, light coupled to a GMR cannot be assigned a propagation direction component along the waveguide. This results in a single sharp dip in the sample's transmission spectrum, caused by waveguide modes with in-plane momentum $\vec{k}_{xy, \text{waveguide}} =\pm m\vec{k}_g$. Non-normal incidence light already carries in-plane momentum, $\vec{k}_{xy, \text{inc}}$, changing the momentum matching condition. The resonance condition can now be fulfilled for waveguide modes of momentum $\vec{k}_{xy, \text{waveguide}} = \vec{k}_{xy, \text{inc}} \pm m\vec{k}_g$. At a given $m$, these two conditions are fulfilled for incident light with different frequencies. This is recognizable as an angle-dependent spectral splitting of the transmission dip, visible in Fig.~\ref{fig:device}e for resonances A and B, observed for rotation of the sample about the $y$-axis. Resonance C would exhibit an analogous splitting for rotation around the $x$-axis, as this frequency couples to guided modes with momentum oriented along the orthogonal lattice direction and therefore requires a component of the wavevector along the orthogonal direction to lift the degeneracy. 

At normal incidence, the symmetry of the GMRs' $E_{x, \omega}$, $E_{z, \omega}$, and $E_{x, 2\omega}$ (illustrated by the field cross-sections in Fig.~\ref{fig:device}b, top row) yields a vanishing overlap integral (Fig.~\ref{fig:device}d). The introduction of a small incidence angle breaks the symmetry (see Fig.~\ref{fig:device}b, bottom row) and creates a large overlap between the required field components (see Fig.~\ref{fig:device}d). The wavelength of maximal overlap closely follows the GMR position (compare Fig. \ref{fig:device}c, d), enabling device tuning. We optimized for a modal overlap peaking at $0.3 ^\circ$ incidence angle (see the spectral cut in Fig. \ref{fig:device}f).

\section*{Metasurface Enhanced Nonlinear Response}
Following the schematic in Fig. \ref{fig:device}a, we fabricated a $600$ $\mu$m $\times$ $600$ $\mu$m large TiO$_2$ metasurface on the surface of the MQW structure. Fabrication details are given in the Methods. 

\begin{figure}[thb]
    \centering
    \includegraphics[width=0.95\textwidth]{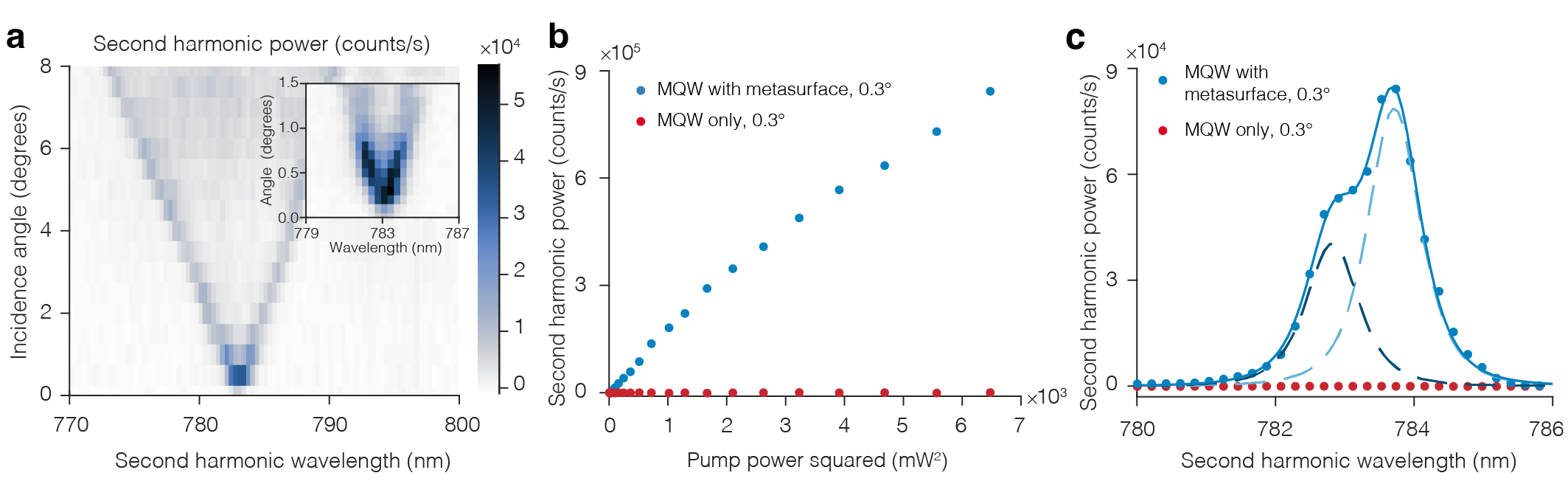}
    \caption{ \textbf{Measured nonlinear device response.}
    \textbf{a} Experimental pump-incidence-angle-resolved second harmonic spectra measured at a pump power of 41 mW. The second harmonic maxima follow the angular evolution of the GMR branches (compare with Figure \ref{fig:device}). The insert shows a high resolution scan from $0^{\circ}$ to $1.5^{\circ}$. At normal incidence, the measurement shows negligible second harmonic response, in accordance with the simulated $\beta$.
    \textbf{b} Measured second harmonic power as a function of the squared average pump power, demonstrating a quadratic dependence at lower pump powers. 
    \textbf{c} Second harmonic spectrum measured at a pump power of 80 mW for the bare MQW and the MQW with a surface patterned metasurface, both at $0.3^{\circ}$ angle of incidence. The 
    second harmonic spectrum generated by the metasurface-MQW-device is fitted with the sum of two squared Lorentzians. Dashed lines show the individual squared Lorentzian fits and the solid line shows the sum. 
    Detector noise has been removed by subtracting a constant background intensity from all measurements.
    }
    \label{fig:sh_measurement}
\end{figure}

We illuminate the sample with $x$-polarized light and rotate it around the $y$-axis (see the Supplementary Information for a schematic of the experimental setup and laser details) and measure the second harmonic response in transmission, shown in Fig. \ref{fig:sh_measurement}. Resonance B, for which the simulations predicted the highest quality factor, generates the highest second harmonic power; thus, we concentrate the following discussion on it. Our measurements show the realized resonance at 1.567 $\mu$m pump wavelength (i.e., it is shifted by 13 nm compared to the simulations, explainable by a less-than 5\% deviation in the heterostructure's thickness or refractive index). Linear transmission measurements, ellipsometry measurements of the heterostructure  refractive index, and nonlinear co- and cross-polarization measurements are provided in the Supplementary Information. 

Fig. \ref{fig:sh_measurement}a shows the experimentally measured second harmonic power created by the metasurface-heterostructure device for incidence angles varying from $0^\circ$ and $8^\circ$. At normal incidence, we measure a negligible second harmonic, consistent with the vanishing overlap observed in the simulations. At finite incidence angles, second harmonic radiation appears at distinct wavelengths. As predicted by the calculated $\beta$, the second harmonic peaks' spectral positions follow the tunable angle splitting of the GMRs at the pump wavelength. We measure the peak second harmonic power at $0.3^{\circ}$ incidence angle, which scales quadratically with the pump power at low intensities (Fig.~\ref{fig:sh_measurement}b), confirming the second-order nature of the process. At high powers, the second harmonic response shows a slight saturation, which we attribute to two-photon absorption competing with second harmonic generation. For comparison, the unpatterned heterostructure shows no measurable response under the same experimental conditions (see Fig. \ref{fig:sh_measurement}b, c). This verifies that the metasurface enables the $\chi^2_{xzx}, \chi^2_{xxz}$-driven second harmonic generation.

Even for the patterned sample, as the incidence angle increases, the oblique incidence introduces a larger $E_z$ to the heterostructure, which is evident from a broad, low-intensity, sum-frequency background emerging gradually in Fig. \ref{fig:sh_measurement}a. At $0.3^{\circ}$ incidence angle, the absence of a broad background confirms that the field components required for second harmonic generation are introduced by the narrow metasurface resonance. Furthermore, each GMR branch of resonance B contributes independently to second harmonic generation, as evident by the lack of a central peak corresponding to sum frequency generation from the combination of photons in the different resonance branches. The total second harmonic response is therefore well described as the sum of two independent resonances. We therefore fit the second harmonic spectrum with two squared Lorentzians (see Fig, \ref{fig:sh_measurement}c), yielding a 0.90-nm second harmonic linewidth and a 1.39-nm pump resonance linewidth at $0.3^{\circ}$ incidence angle. This corresponds to a quality factor of $1124$, and confirms the high quality of the resonance.

To evaluate the effective nonlinear tensor element of our metasurface-coupled heterostructure, we compare its experimentally measured second harmonic flux (at $0.3^{\circ}$ incidence angle) with that generated by the unpatterned heterostructure (at $45^{\circ}$ incidence angle). We reproduce the data by modeling second harmonic and sum frequency generation for both measurements, taking into account the heterostructure's refractive index, thin-film interference, reflections, and the metasurface's spectral acceptance versus the pump laser's bandwidth (which is $\sim 100\times$ larger than the GMR bandwidth, see Methods).

At 1.57 $\mu$m, we find that the metasurface enhances the pump's $E_xE_z$ field product in the heterostructure by $\sim57\times$ compared to the $E_xE_z$ field product present in the heterostructure at $45^{\circ}$ incidence. Furthermore, our model shows a $\sim 11.5\times$ enhancement when comparing the metasurface-generated $E_xE_z$ field product in the heterostructure compared to the $E_xE_x$ field product present in the heterostructure at normal incidence (which does not require polarization conversion).

As the metasurface allows for the sample to be used with $x$-polarized light at normal incidence, this yields an effective, usable $\chi^{(2)}_{\mathrm{eff,MQW+MS}} = \chi^{(2)}_{xxx\mathrm{,MQW+MS}} \approx 14$ nm/V; more than 270 times larger than the largest $\chi^{(2)}$ tensor element of LiNbO$_3$.  Note that correcting for absorption of the second harmonic radiation in the heterostructure, one obtains a nonlinearity of 17 nm/V. Here, we report the lower nonlinear susceptibility, as it represents the useful output from the device.

Optimal use of any resonance-enhanced device requires matching the resonance bandwidth to the pump source, as all spectral content outside of the resonance bandwidth is discarded. Here, the MQW design allows us to spectrally tune the material resonance, and the spectral position and linewidth of the GMR can be altered by optimizing the metasurface design. This enables closed-loop design and optimization of nonlinear devices where every aspect of the polarization is tailored to the intended application.

\section*{Conclusions}
By coupling interband-engineered MQWs to high quality factor metasurface resonances, we demonstrate a co-design approach that simultaneously enhances the intrinsic nonlinear susceptibility and provides optical access to non-trivial tensor elements at near-infrared wavelengths. The GaAs/AlGaAs heterostructure grown here exhibits a $\chi^{(2)}_{\mathrm{eff,MQW}}$ tensor element more than thrice that of bulk GaAs; the metasurface achieves an additional factor-ten increase. Even larger nonlinear susceptibilities can be achieved by optimizing fabrication and introducing additional degrees of freedom to the heterostructure design.

The versatility of the metasurface can extend our demonstration to a range of applications. While we used a nonlocal metasurface for field enhancement and polarization conversion, other metasurfaces can spatially engineer the nonlinear output by controlling the propagation phase or geometric phase. Beyond second harmonic generation, this concept can also be utilized for difference frequency generation, and can enable entangled photon generation through spontaneous parametric down-conversion. As the device is entirely based on dielectric and semiconducting materials, it combines a high damage threshold and foundry compatibility. The considerable $\chi^{(2)}$ magnitude increase with respect to the bulk $\chi^{(3)}$ enables devices with less three-photon absorption, hence allowing for further miniaturization or operation under relaxed phase matching conditions compared to their bulk crystal counterparts. 

The combination of plasmonic metasurfaces and quantum wells has proven powerful in the mid-infrared \cite{Lee2014}, and the present demonstration establishes the near-infrared to visible counterpart. Furthermore, the presented approach to band engineering can be extended to higher-order nonlinearities, enabling Kerr-based nonlinear devices with tailored nonlinear susceptibilities. 
The work highlights the power of uniting bandstructure engineering with resonant nanophotonics and opens doors for high-efficiency devices at wavelengths of great technological importance. 
\newpage

\section*{Methods} \label{Methods}

\subsection*{Quantum mechanical simulation and $\chi^{(2)}$ calculation}
Starting from the general form of the dipole matrix formalism for the second-order nonlinear susceptibility 
\cite{boyd_nonlinear_2019, bloembergen_nonlinear_1996, khurgin_secondorder_1987}, the expression for the electron ($e$) and the heavy hole ($hh$) susceptibility of the resonant tensor element can be expressed as

\begin{equation}
    \chi_{xzx,e}^{(2)}(\omega_1 + \omega_2, \omega_1,\omega_2) = \frac{N_z e^3 r_{e,hh}^2}{6\epsilon_0 \hbar^2}\sum_{k_{||}}\sum_{m,n}\sum_{l}\Biggl(\frac{\langle\psi_{hh,m}|\psi_{e,n}\rangle\langle\psi_{e,n}|z|\psi_{e,l}\rangle\langle\psi_{e,l}|\psi_{hh,m}\rangle}{(\omega_{hh,m}^{e,n}(k_{||})-\omega_1-\omega_2+i\Gamma)(\omega_{hh,m}^{e,l}(k_{||})-\omega_1+i\Gamma)} \Biggl)
    \label{eqn:chi2equation_electron}
\end{equation}

\begin{equation}
            \chi_{xzx,hh}^{(2)}(\omega_1 + \omega_2, \omega_1,\omega_2) = \frac{N_z e^3 r_{e,hh}^2}{6\epsilon_0 \hbar^2}\sum_{k_{||}}\sum_{m,n}\sum_{l}\Biggl( -  \frac{\langle\psi_{e,n}|\psi_{hh,m}\rangle\langle\psi_{hh,m}|z|\psi_{hh,l}\rangle\langle\psi_{hh,l}|\psi_{e,n}\rangle}{(\omega_{hh,m}^{e,n}(k_{||})-\omega_1-\omega_2+i\Gamma)(\omega_{hh,l}^{e,n}(k_{||})-\omega_1+i\Gamma)}\Biggl)
    \label{eqn:chi2equation_hole}
\end{equation}

where $[m,n,l]$ denote bound states in the conduction and heavy-hole bands, $\psi_{e,hh}$ are the corresponding envelope wavefunctions, $\omega_{1,2}$ denote the input photon frequencies, $\omega_{hh;m,l}^{e;n,l}$ represent interband transition energies, $\Gamma$ is the phenomenological broadening parameter (here set to 5 meV), $k_{||}$ denotes the in-plane momentum, $N_z$ is the spin degeneracy, and $r_{e,hh}=\langle u_e|r|u_{hh}\rangle$ is the interband dipole matrix element of a Bloch function. The electron and heavy hole susceptibilities contain the intersubband matrix elements of the electrons and heavy holes, respectively.

The envelope functions of the coupled quantum wells were obtained from self-consistent Schrödinger–Poisson simulations performed with Nextnano \cite{birner_nextnano_2007}. The interband matrix element $r_{e,hh}$ for GaAs was calculated using density functional theory in VASP with the HSE06 hybrid functional. The summation over in-plane $k$ states was evaluated by converting it to a two-dimensional integral over $(k_x,k_y)$, truncated at one-tenth of the Brillouin zone, beyond which the contribution to $\chi^{(2)}$ was found to be negligible \cite{Ramesh2023}.

\subsection*{Material growth and substrate transfer}

The III-V heterostructures were grown on semi-insulating GaAs (100) wafers using molecular beam epitaxy in a Varian Gen II system. The system is equipped with solid-source thermal effusion cells for Al and Ga and a solid-source valved cracker for As. The growth temperature was maintained at 600°C, monitored by band edge thermometry. AlGaAs was grown at 1.85 \r{A} s$^{-1}$ and GaAs was grown at 0.83 Å s$^{-1}$ under a 15$\times$ As overpressure. The heterostructure contains 16 periods each comprised of AlGaAs barrier layers (shown in bold) and two GaAs quantum wells, following the layer sequence (in nanometers) \textbf{18.2}/7.1/\textbf{1.8}/2.9. Etch stop layers were grown to enable subsequent flip-chip transfer to Al$_2$O$_3$ substrates. The substrate transfer was done by bonding the heterostructure to Al$_2$O$_3$ with 353ND, EPO-TEK epoxy resin, followed by mechanical lapping for etching of the GaAs substrate and wet etch removal of the etch-stop layers (see schematic in the Supplementary Information.).

\subsection*{Electron microscopy characterization of MQW film}
Using a Focused Ion Beam (FIB), lamellae for scanning transmission electron microscopy (STEM) were prepared, and subsequently fine-polished with Ar using a nano-mill. Using a double aberration corrected JEOL ARM-300CF microscope operating at 300 kV, STEM images were acquired. We performed the z-contrast high-angle annular dark-field (HAADF-STEM) imaging with a probe convergence semi-angle of 25.7 mrad and an inner collection angle of 53 rad. 

EDS maps were taken with dual 100 mm$^2$ Si drift detectors. The EDS maps were constructed by summing 100 drift-corrected scans, each with a dwell time and step size of 10 ms and 52 pm, respectively. Subsequently, the composition profiles of Ga and Al were fitted as piecewise linear functions in the growth direction to be used for the quantum mechanical simulations of the bound state wavefunctions and energy levels of the grown material. 

\subsection*{Metasurface simulations}
The electromagnetic simulations were carried out with rigorous coupled wave analysis using GRCWA \cite{jin_inverse_2020}. The structural parameters used in the simulations were a pillar height of 390 nm, a radius of 230 nm, an $x$ periodicity of 891 nm, and a $y$ periodicity of 650 nm, with p-polarized incident light. The refractive index of the TiO$_2$ was measured by ellipsometry of an amorphous TiO$_2$ thin-film deposited by ALD under the same conditions as the pillars of the metasurface. The refractive index of the MQW stack was determined by ellipsometry of the equivalent GaAs/AlGaAs stack on a GaAs substrate. In the simulations, the MQW stack was treated as a single material layer with the measured refractive index. Equivalent simulations were completed for the pump and second harmonic wavelengths, and the corresponding pairs were used to compute the modal overlap. 

\subsection*{Metasurface fabrication}
Undiluted positive-tone electron beam lithography resist, ZEP 520A, was spin-coated at 3800 rpm for 45 s. The sample was pre-baked on a hotplate at 90 $^{\circ}$C for 3 minutes, followed by a hotplate at 180 $^{\circ}$C for 3 minutes. A conductive polymer (Showka Denko ESPACER 300) was spun at 1500 rpm for 45 s to avoid charging effects. The resist was patterned using electron beam lithography (Elionix BODEN 150) with an acceleration voltage of 150 kV and a current of 1 nA. The sample was developed with o-Xylene under gentle agitation. TiO$_2$ was deposited by ALD (Savannah, Cambridge Nanotech), followed by reactive ion etching (Oxford PlasmaPro 100 Cobra 300) of overgrown TiO$_2$. Finally, the resist was removed with Remover PG. See schematic in the Supplementary Information.

\subsection*{Experimental characterization}
Linear optical characterization of the unpatterned GaAs/AlGaAs heterostructure was done using a Cary 7000 Universal Measurement Spectrophotometer. Linear and nonlinear characterization of the metasurface device was done using a MenloSystems ELMO Femtosecond Erbium Laser centered at 1560 nm with an average output power of 330 mW, pulse duration of 70 fs, and repetition rate of 100 MHz. The linear signal was collected with an Anritsu MA9710B optical spectrum analyzer, and the nonlinear response was measured with an Andor SR-500i-B2-R spectrometer. The optical characterization setup is shown in detail in the Supplementary Information. 

\subsection*{Measurement of $\chi^{(2)}_{xzx}(\omega)+\chi^{(2)}_{xxz}(\omega)$ by comparison with LiNbO$_3$}
To determine the magnitude of $\chi^{(2)}_{xzx}(\omega)+\chi^{(2)}_{xxz}(\omega)$ in the MQW, we performed relative measurements against a reference sample. A 0.60 $\mu$m thick x-cut congruent thin-film of LiNbO$_3$ on sapphire was measured at normal incidence, and a 0.60 $\mu$m thick MQW film on sapphire substrate was measured at $45^\circ$ incidence. The samples were excited from the substrate side under equal pumping conditions, and the transmitted pump and generated second harmonic spectra were recorded.
To our knowledge, literature reports no congruent LiNbO$_3$ second-order nonlinear susceptibility $\chi^{(2)}_{zzz}(1567 \text{ nm}) = 2d_{33}$ measured close to 1567 nm pump wavelength.
The most cited measured value is $\chi^{(2)}_{zzz}(1064 \text{ nm}) = 54.4$ pm/V \cite{miller_dependence_1971, nikogosyan_nonlinear_2005, zhu_integrated_2021}.

Applying Miller's rule \cite{miller_optical_1964} to this value predicts $\chi^{(2)}_{zzz}(1567 \text{ nm}) = 51.9$ pm/V. 
Careful measurements by Shoji et al. found $\chi^{(2)}_{zzz}(852 \text{ nm}) = 51.4$ pm/V, $\chi^{(2)}_{zzz}(1064 \text{ nm}) = 50.4$ pm/V, and $\chi^{(2)}_{zzz}(1313 \text{ nm}) = 39.0$ pm/V (\cite{shoji_absolute_1997}); however, the latter value deviates considerably from the Miller's rule expectation.
The highest $\chi^{(2)}_{zzz}(1064 \text{ nm}) = 83.4$ pm/V was reported for stoichiometric LiNbO$_3$, which contains a higher lithium fraction than congruent LiNbO$_3$ \cite{roberts_simplified_1992, nikogosyan_nonlinear_2005}. For consistency with the current body of literature, this work uses $\chi^{(2)}_{zzz}(1567 \text{ nm}) = 51.9$ pm/V.

To account for reflections at the sample interfaces and thin film interference, we model propagation into our experimental structures, including the respective incidence angle, using finite difference time domain simulations (Flexcompute Tidy3D), see \url{https://github.com/marcus-o/linbo_mqw_comparison}. To model the experimental pump pulse spectrum, we Fourier transform the incident laser pulses' spectrum and inject the resulting electric field using a time-dependent source and record the polarization-resolved time-dependent electric field $E_x(t)$ and $E_z(t)$ at the center position of our thin films. 

Whereas the frequency dependence of the nonlinear response of LiNbO$_3$ is small in our laser's wavelength range, we account for the heterostructure's frequency dependence by modeling $\chi^{(2)}$ in the time domain as the product of two independent and exponentially decaying (lifetime or dephasing time: 30 fs) oscillators. The coherence lengths for second harmonic generation from 1550 nm to 775 nm wavelength are 1.8 $\mu$m in the heterostructure and 9.6 $\mu$m in LiNbO$_3$. As the samples are considerably thinner, phase matching and pump depletion are negligible, and we assume sum-frequency radiation builds up coherently along the sample. We correct for the sum-frequency propagation  and absorption ($n_{780 \text{ nm}} = 3.39 + 0.11i$ from ellipsometry measurements) outside of the samples using a finite difference time domain simulation. Under the assumption that the laser pulses' temporal/spectral phases do not possess strong second-order (chirp) or higher-order phase components in the samples, the simulated and experimentally measured sum-frequency spectra match (see Figure in Supplementary Information) \cite{miranda_simultaneous_2012}.

Finally, we scale the heterostructure's $\chi^{(2)}_{xzx}(\omega)+\chi^{(2)}_{xxz}(\omega)$ so the simulated and experimentally measured sum-frequency fluxes match for LiNbO$_3$ and the heterostructure. This relative approach removes the need for an absolute pump intensity calibration and enables a direct comparison between the MQW and LiNbO$_3$ under consistent experimental and numerical conditions.

\subsection*{Measurement of resonant second harmonic generation enhancement}
To measure the resonant enhancement by the GMR, we compare the second harmonic flux generated by the metasurface sample at a $0.3^\circ$ incidence angle with the sum-frequency generated by the bare MQW film at $45^\circ$ incidence angle using otherwise identical excitation conditions. 

Experimentally, for the metasurface sample at $0.3^\circ$ incidence angle, we observe no sum-frequency radiation generated by mixing pump photons from different branches of the GMR (which would be observable as additional peaks in the sum-frequency spectrum, see Fig.~\ref{fig:sh_measurement}a) and no sum-frequency radiation generated by mixing one GMR photon with a non-resonant photon (which would be observable as a broad sum-frequency background, see Fig.~\ref{fig:sh_measurement}a). Therefore, we fit the generated second-harmonic radiation using the sum of two squared Lorentzian profiles (see Fig.~\ref{fig:sh_measurement}c) to extract the resonance frequencies and widths of the GMR branches. 

We account for the effects of the resonant metasurface in the simulation procedure detailed above by applying a Lorentzian filter to the electric field in the nonlinear layer before calculating sum-frequency generation. The filter uses the experimentally determined center frequency and width. We then scale the $E_x(t) E_z(t)$ product in the heterostructure until the simulated and experimentally measured sum-frequency spectra match those of the metasurface-covered and bare heterostructure, removing the need for an absolute pump intensity calibration.

\newpage
\subsection*{Acknowledgements}

The authors thank Marko Lon\v{c}ar at Harvard University for providing access to the MenloSystems ELMO Femtosecond Erbium Laser, and Anna Shelton for assistance with its operation. The authors also thank Joon-Suh Park for helpful discussions. The simulation work was in part performed using the Tidy3D software from Flexcompute.

\subsection*{Funding}
This research was supported by a Multidisciplinary University Research Initiative from the Air Force Office of Scientific Research (AFOSR Award No. FA9550-22-1-0307).
This work was performed in part at the Harvard University Center for Nanoscale Systems (CNS); a member of the National Nanotechnology Coordinated Infrastructure Network (NNCI), which is supported by the National Science Foundation under NSF award no. ECCS-2025158.
P.U.F. thanks the Aker Scholarship
I.O. acknowledges funding from the ZKS Master Thesis Grant.
M.O. acknowledges funding from the European Union (grant agreement 101076933 EUVORAM). The views and opinions expressed are, however, those of the author(s) only and do not necessarily reflect those of the European Union or the European Research Council Executive Agency. Neither the European Union nor the granting authority can be held responsible for them.
IB and HJ acknowledge support by the U.S. Department of Energy (DOE), Office of Science, Basic Energy Sciences (BES), under Field Work Proposal Number 24–017574. This work was performed, in part, at the Center for Integrated Nanotechnologies, an Office of Science User Facility operated for the U.S. DOE Office of Science. Sandia National Laboratories is a multi-mission laboratory managed and operated by National Technology and Engineering Solutions of Sandia, LLC., a wholly owned subsidiary of Honeywell International, Inc., for the U.S. DOE’s National Nuclear Security Administration under contract DE-NA0003525. This paper describes objective technical results and analysis. Any subjective views or opinions that might be expressed in the paper do not necessarily represent the views of the U.S. DOE or the United States Government.

\subsection*{Competing interests}
The authors declare no competing interests.

\subsection*{Data availability }
The data supporting the findings of this study are available at ZENODO.

\subsection*{Code availability }
Exemplary data analysis code used in this study is available at \url{https://github.com/marcus-o/linbo_mqw_comparison}.

\subsection*{Author contributions}
Concept: PUF, MO, SB, FC. 
MQW theory, design, and fabrication: PD, AR, RR, SB. 
MQW substrate transfer: HJ, IB. 
Metasurface design: PUF, MO. 
Device fabrication and characterization: PUF. 
Optical measurements: IO, PUF, TPL, CMS. 
Data analysis: PUF, MO. 
Transmission electron microscopy: YJ, MW, XP. 
Initial draft: PUF, MO. 
All authors read, edited, and agreed with the final manuscript.

\newpage
\printbibliography
\newpage

\newgeometry{left=3.1cm, bottom=3cm, right=3.1cm, top=3cm} 
\setlength{\parindent}{0cm}

\begin{center}
    {\Large \textbf{Supplementary Information} \par \bigskip}
    {\Large \textbf{for } \par \bigskip}
    {\Large \textbf{Quantum-Well-Metasurface to Maximize Nonlinear Polarization} \par \bigskip}

    Pernille Undrum Fathi\textsuperscript{1, *},
    Irene Occhiodori\textsuperscript{1, 2},
    Patrick Devaney\textsuperscript{3},
    Amberly Ricks\textsuperscript{3},\\
    Rithvik Ramesh\textsuperscript{3, 4},
    Yiwei Ju\textsuperscript{5},
    Moaz Waqar\textsuperscript{5},
    Theodore P. Letsou\textsuperscript{1, 6},\\
    Christina M. Spägele\textsuperscript{1},
    Hyunseung Jung\textsuperscript{7},
    Igal Brener\textsuperscript{7},
    Xiaoqing Pan\textsuperscript{5, 8, 9},\\
    Marcus Ossiander\textsuperscript{1, 10, *},
    Seth R. Bank\textsuperscript{3},
    Federico Capasso\textsuperscript{1, *}
    \bigskip
\end{center}

{\footnotesize
\textsuperscript{1} Harvard John A. Paulson School of Engineering and Applied Sciences, Harvard University, Cambridge, 02138, Massachusets, USA

\textsuperscript{2} Department of Materials, ETH Zürich, Zürich, 8093, Switzerland

\textsuperscript{3} Microelectronics Research Center and Electrical and Computer Engineering Department,The University of Texas at Austin, Austin, 78712, Texas, USA

\textsuperscript{4} Department of Electrical Engineering, California Institute of Technology, Pasadena, 91125, California, USA

\textsuperscript{5} Department of Materials Science and Engineering, University of California, Irvine,  Irvine, 92697, California, USA

\textsuperscript{6} Department of Electrical Engineering and Computer Science, Massachusetts Institute of Technology, Cambridge, 02142, Massachusets, USA

\textsuperscript{7} Center for Integrated Nanotechnologies, Sandia National Laboratories, Albuquerque, 87185, New Mexico, USA

\textsuperscript{8} Department of Physics and Astronomy, University of California, Irvine,  Irvine, 92697, California, USA

\textsuperscript{9} Irvine Materials Research Institute (IMRI), University of California, Irvine,  Irvine, 92697, California, USA

\textsuperscript{10} Institute of Experimental Physics, Graz University of Technology, Graz, 8010, Austria}
\bigskip

* pernillefathi@g.harvard.edu, marcus.ossiander@tugraz.at, capasso@seas.harvard.edu

\restoregeometry 
\setlength{\parindent}{0.5cm}
\newpage

\begin{figure}[ht]
    \centering
    \includegraphics[width=1\linewidth]{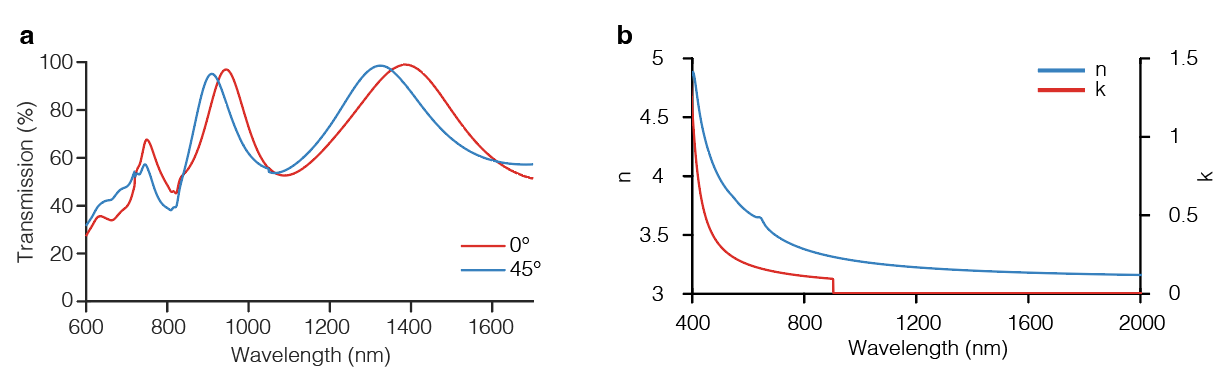}
    \caption{ \textbf{Characterization of the substrate transferred heterostructure.}
    \textbf{a} Transmission spectrum of a 16-period MQW film on sapphire measured at incidence angles of $0^{\circ}$ and $45^{\circ}$ using a Cary 7000 UV-Vis-NIR Universal Measurement Spectrophotometer. Linear transmission measurements of the grown sample at normal and $45^{\circ}$ incidence show only above-band gap absorption and thin-film Fabry Perot interference. Grating change at 720 nm, detector change at 1050 nm.
    \textbf{b} Refractive index measurements of the heterostructure on a sapphire substrate taken using a Woollam RC2 Spectroscopic Ellipsometer.
    }
    \label{fig:lin_characterization_material}
\end{figure}

\begin{figure}[ht]
    \centering
    \includegraphics[width=1\linewidth]{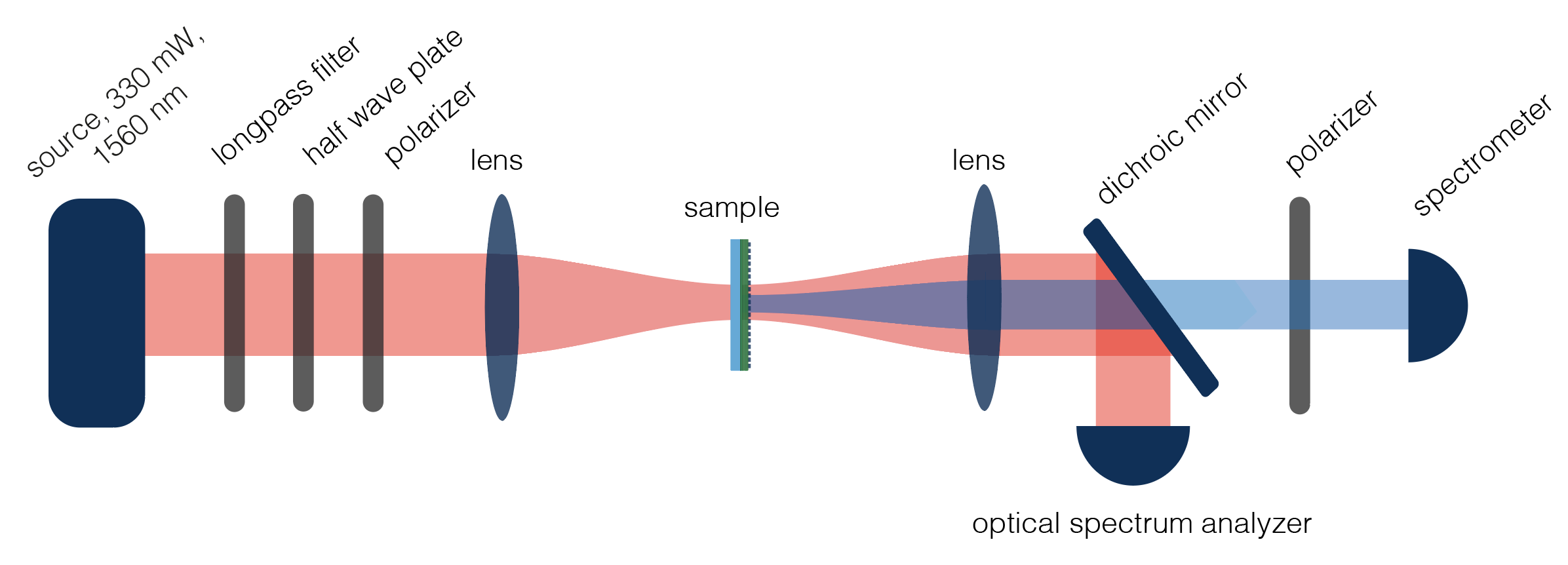}
    \caption{ \textbf{Schematic of the optical setup for device characterization.}
    70 fs pulsed light from a MenloSystems ELMO Femtosecond Erbium laser passes through a long pass filter to remove second harmonic light coming from the laser, followed by a half-waveplate and a polarizer for polarization control. A low numerical aperture lens with a focal length of 40 cm focuses light onto the sample. The light is collected by a 50 cm focal length lens, followed by a dichroic mirror to block the pump frequency. A second polarizer enables measurements of the output polarization, and the output is detected using an Andor SR-500i-B2-R spectrometer. The pump frequency is collected using an Anritsu MA9710B optical spectrum analyzer. 
    }
    \label{fig:setup}
\end{figure}

\begin{figure}
    \centering
    \includegraphics[width=0.65\linewidth]{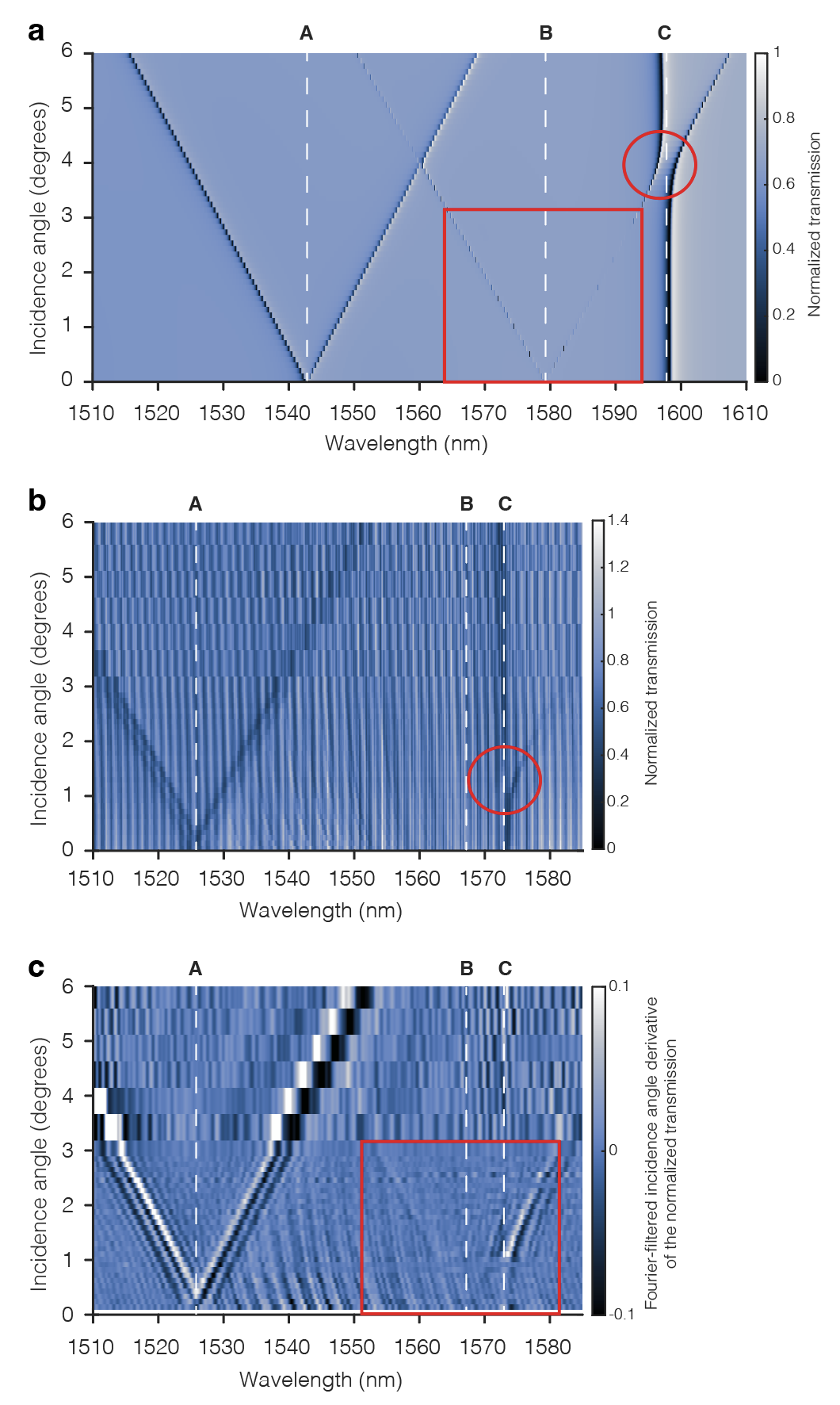}
    \caption{\textbf{Incidence angle-resolved transmission spectra of the metasurface for x-polarized illumination. }
    \textbf{a} Simulated results. 
    \textbf{b} Experimental measurements. The high quality factor resonance B is most visible in the anti-crossing with resonance C, marked by a red circle in panels \textbf{a} and \textbf{b}. Note that a relative wavelength shift between resonances B and C leads to the anti-crossing occurring at a smaller angle in the experimentally measured spectra compared to the simulations.
    \textbf{c} Fourier-filtered incidence angle derivative of the experimental measurements. Here, resonance B is visible, marked by a red square in panels \textbf{a} and \textbf{c}.
    }
    \label{fig:metasurface_transmission}
\end{figure}

\begin{figure}[ht]
    \centering
    \includegraphics[width=1\linewidth]{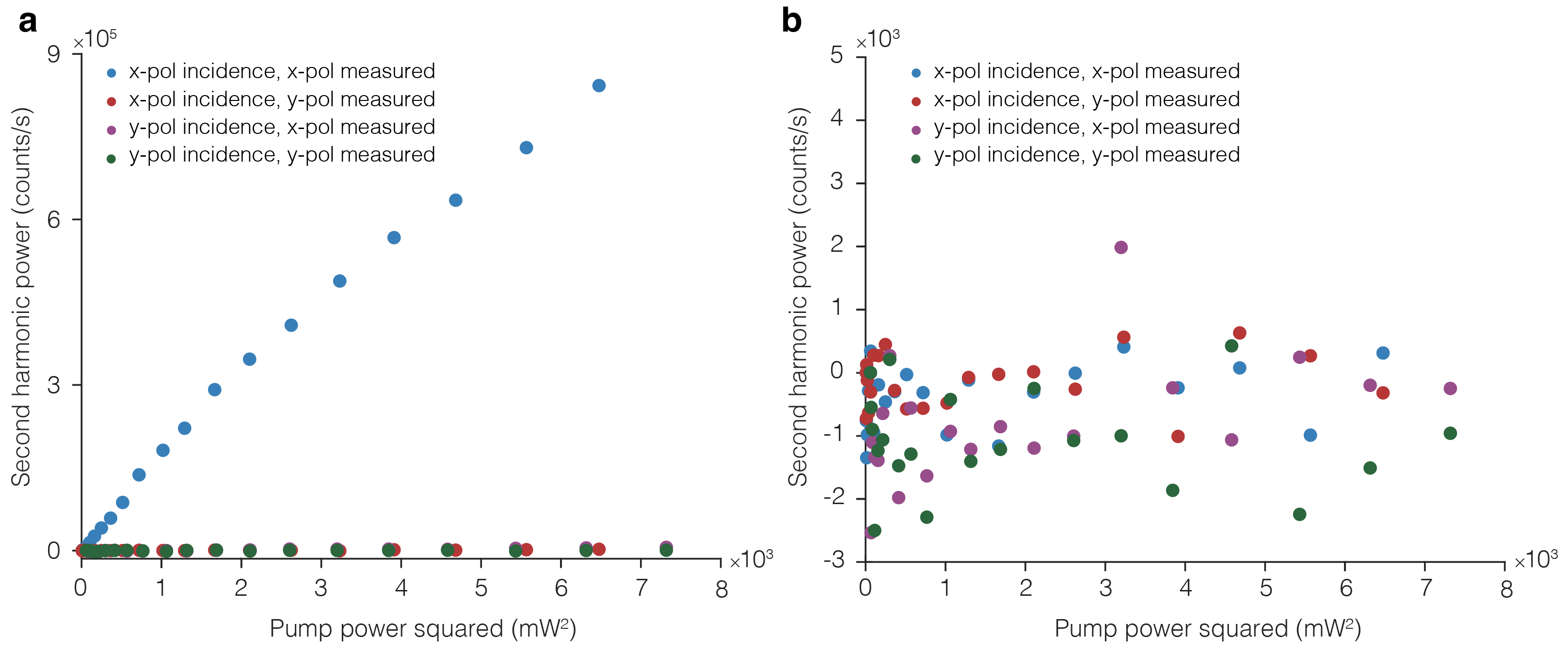}
    \caption{\textbf{Nonlinear response of the metasurface/heterostructure sample and the bare heterostructure. } 
    Measured generated second harmonic power as a function of the squared average incident pump power and the input and output polarization (linear $x$ and $y$ polarized light) for \textbf{a} the metasurface/heterostructure sample, and \textbf{b} the bare heterostructure, both measured at $0.3^{\circ}$ angle of incidence.
    }
    \label{fig:co_cross_power_sweeps}
\end{figure}

\begin{figure}[ht]
    \centering
    \includegraphics[width=1\linewidth]{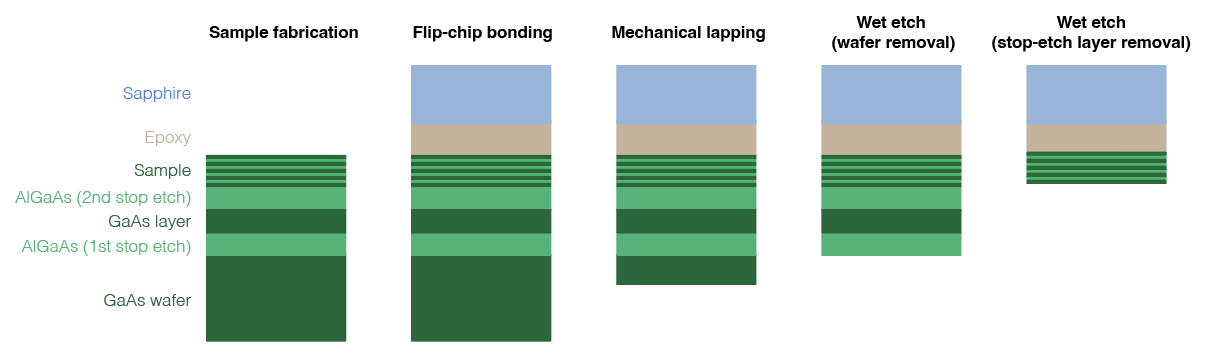}
    \caption{\textbf{Substrate transfer.}
    The sample is grown via molecular beam epitaxy and subsequently flip-chip bonded to a 500-$\mu$m thick sapphire substrate using 353ND EPO-TEK epoxy resin. Mechanical lapping is used for the first stage of the GaAs substrate removal, followed by a wet etch removal of the remaining GaAs substrate and the etch-stop layers. 
    }
    \label{fig:substrate_transfer}
\end{figure}

\begin{figure}[ht]
    \centering
    \includegraphics[width=1\linewidth]{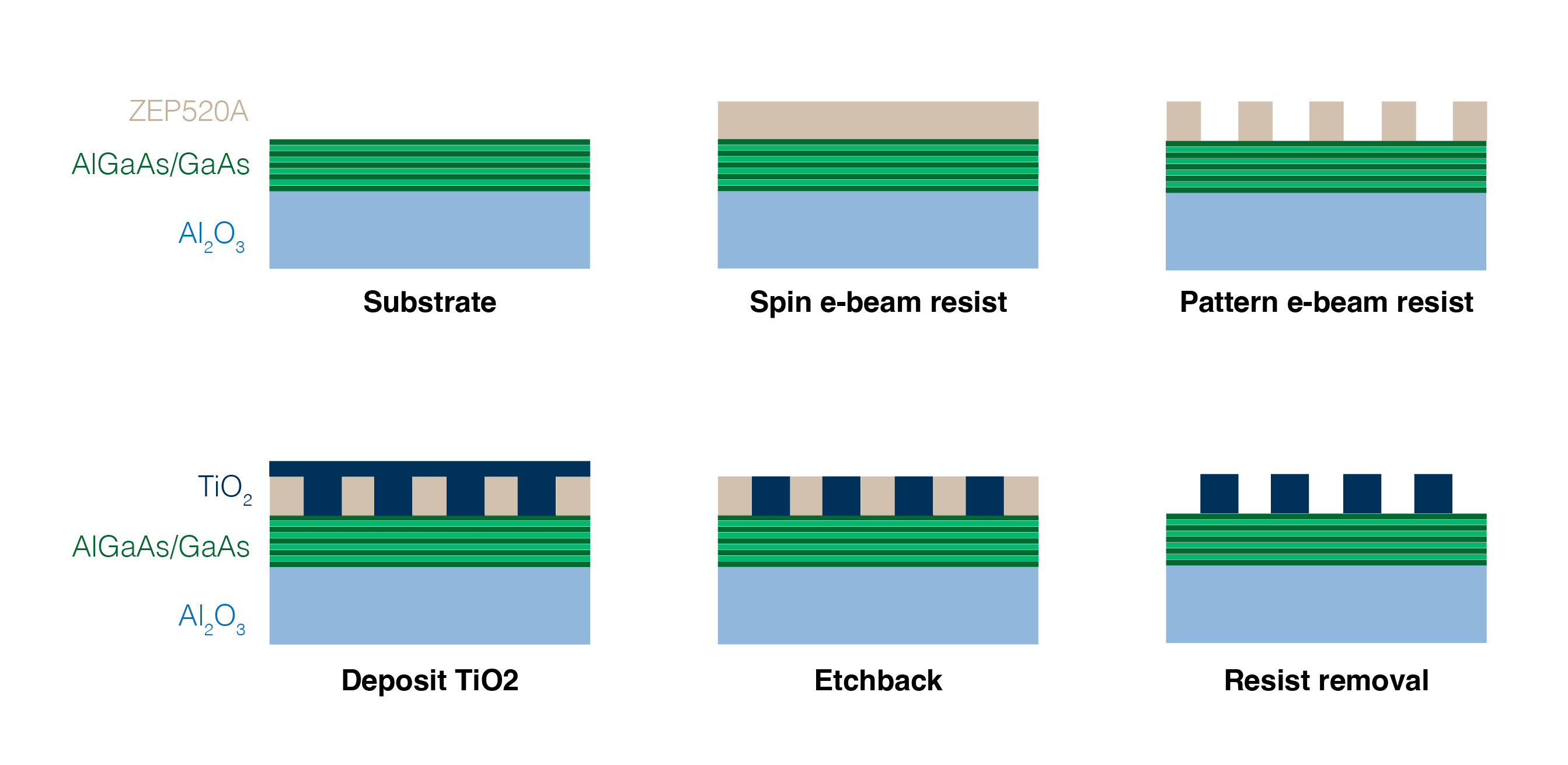}
    \caption{\textbf{Metasurface fabrication.} Positive-tone E-Beam resist is spin coated on the surface of the heterostructure and exposed using electron beam lithography. The exposed resist is removed using developer and the gaps are filled with TiO$_2$ by atomic layer deposition. The overgrown TiO$_2$ is removed by reactive ion etching, and the remaining resist is removed.
    }
    \label{fig:metasurface_fabrication}
\end{figure}

\begin{figure}[ht]
    \centering
    \includegraphics[width=0.5\linewidth]{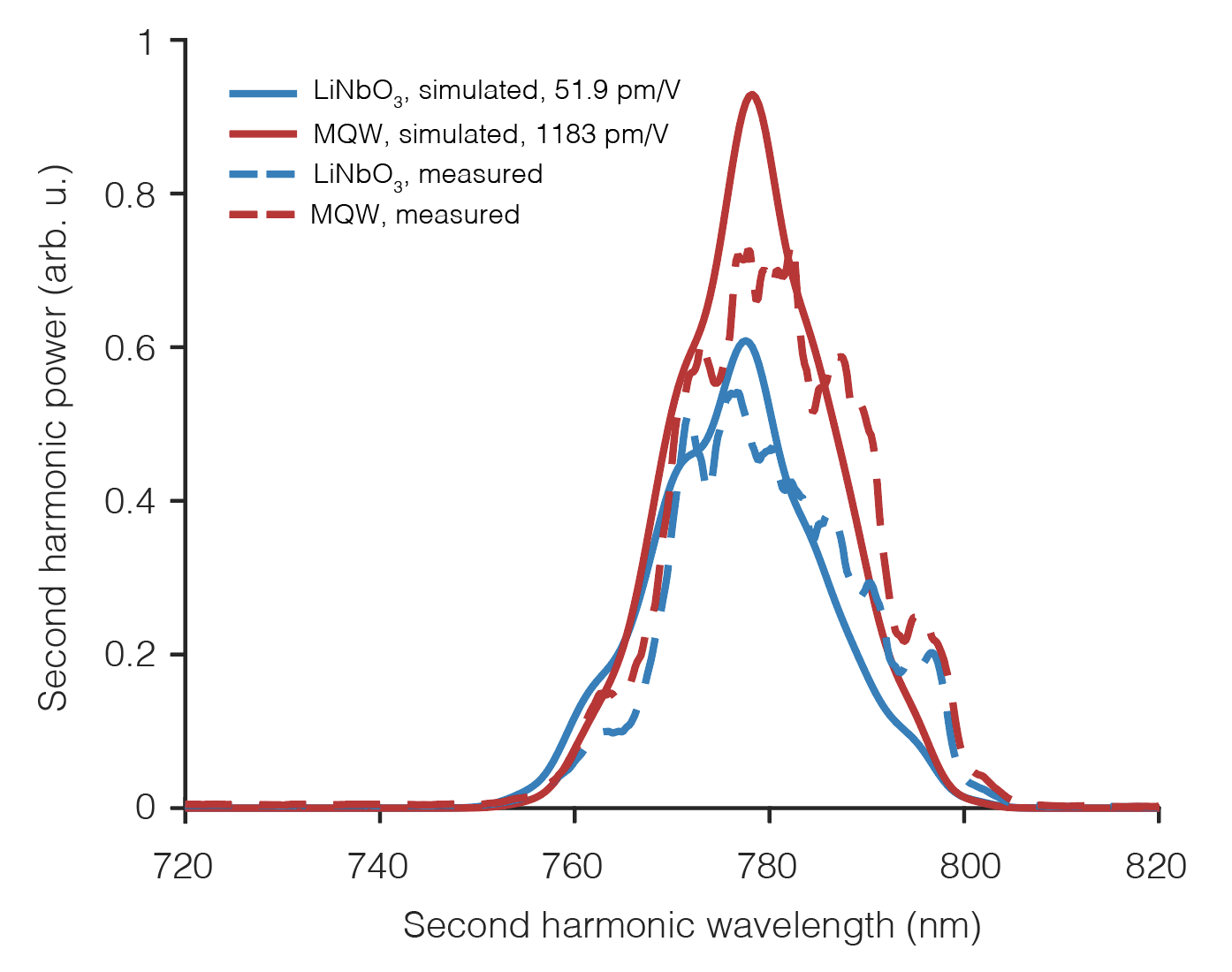}
    \caption{\textbf{Measured and modeled sum-frequency spectra.} The output sum-frequency spectra of a 16-period MQW on sapphire measured at $45^{\circ}$ incidence and x-cut LiNbO$_3$ measured at $0^{\circ}$ incidence under x-polarized illumination. Dashed lines indicate experimental measurements, while solid lines show simulated sum-frequency spectra derived from the measured incident spectrum. Both blue and red curves yield the same integrated power, respectively.
    }
    \label{fig:SFG}
\end{figure}

\end{document}